\documentstyle[preprint,aps]{revtex}
\tighten
\input psfig.tex
\begin{document}
\draft

\title{Transfer-matrix scaling from disorder-averaged correlation lengths\\ for diluted Ising systems}
\author{S. L. A. de Queiroz$^a$\cite{email} and R. B. Stinchcombe$^b$}
\address{
$^a$Departamento de F\'\i sica, Pontif\'\i cia Universidade Cat\'olica do Rio de Janeiro,\\ Cx.  Postal 38071, 22452-970 Rio de Janeiro, RJ, Brazil\\
$^b$ Department of Physics, Theoretical Physics, University of Oxford,\\
 1 Keble Road, Oxford OX1 3NP, United Kingdom}
\date{\today}
\maketitle
\begin{abstract}
A transfer matrix scaling technique is developed for
randomly diluted systems, and applied to the site-diluted Ising
model on a square lattice in two dimensions. For each allowed disorder configuration between two adjacent columns, the contribution of the respective transfer matrix to the decay of correlations is considered only as far as the ratio of its two largest eigenvalues, allowing an economical calculation of a configuration-averaged correlation length. Standard phenomenological-renormalisation procedures are  then used to analyse aspects of the phase boundary which are difficult to assess
accurately by alternative methods. For magnetic site concentration
$p$ close to $p_c$, the extent of exponential behaviour of the $T_c \times p$ curve is clearly seen for over two decades of variation of $p - p_c$.
Close to the pure-system limit, the exactly-known reduced slope is
reproduced to a very good approximation, though with non-monotonic convergence.
The averaged correlation lengths are inserted into the exponent-amplitude relationship predicted by conformal invariance to hold at criticality. The resulting exponent $\eta$ remains near the pure value (1/4)
for all intermediate concentrations until it crosses over to the
percolation value at the threshold. 
\end{abstract}

\narrowtext
\newpage
The transfer matrix is well known to provide exact solutions for 
a number of low-dimensional pure systems such as spin models etc~\cite{baxter}. 
Applied to finite-width strips and combined with finite size
scaling~\cite{fisher}, it has also proved to be extremely powerful and
accurate for non-solvable pure cases, ranging from magnetic
systems~\cite{fsgen} to walks~\cite{dersaw} and quantum
Hamiltonians~\cite{malte}.
 The application of such
``strip-scaling'' techniques to random systems has however been
extremely limited, despite the enormous interest in such problems as
the spin glass, random field and dilute magnets, ceramic
superconductors etc. This is because the form so far
utilized~\cite{cheung,huse,glaus1,glaus2} applies the strip scaling to 
particular realisations of the
random system which need not be representative unless appropriate
averaging (or self-averaging) is employed, resulting in very
large scale computing on extremely long strips and/or many realisations.

Here, we provide a strip scaling
approach for random systems, in which the disorder averaging is
carried out as one proceeds along the strip, and
which therefore does not have the deficiencies noted above. Results reported here extend, and give details of, those contained in a previous Rapid Communication\cite{dqrbs}.
Our scheme does rely on certain assumptions regarding the dominant
contributions to the decay of correlations, which must be spelt out clearly
and supported by numerical evidence whenever possible. This is one of our
purposes in what follows.

The main numerical application of the technique here is to the two-dimensional site-diluted Ising model\cite{stinch}. The transfer-matrix descriptions of the
special limiting cases of percolation\cite{derrida} and pure Ising system\cite{night} are well-understood, and are reproduced by the present approach, as we shall see below. On the other hand, several aspects of the behaviour at intermediate dilution are either not known to the same degree of
accuracy, or have not been successfully accounted for via transfer-matrix 
methods. Of the former, we shall focus on the extent of exponential behaviour of the $T_c \times p$ curve for magnetic site concentration $p$ close to $p_c$, and
on the variation of the decay-of-correlations exponent $\eta$ along the critical line; as for the latter, we give results for the reduced slope of the phase boundary close to the pure-system limit, to be compared with an exact solution
by Thorpe and McGurn\cite{thmg}.

In the transfer-matrix scaling formulation,  the crucial quantity to be considered is the rate of decay of correlations. On a strip, the correlation
functions always (i.e. at finite temperatures) decay exponentially for suitably large distances:
\begin{equation}
<\sigma_0 \sigma_R> \sim \exp( -R/\xi) \ ,
\label{eq:1}
\end{equation}
\noindent with the correlation length $\xi$ being given, for a pure system, by
\begin{equation}
\xi^{-1}  = -\ln {\lambda_{2}}/{\lambda_{1}} \ ,
\label{eq:2}
\end{equation}
\noindent where $\lambda_{1}, \lambda_{2}$ are 
the two largest eigenvalues of
the transfer matrix~\cite{domb}. To obtain this pure system result, the effect of repeated applications of the (same) transfer matrix is investigated,
and only leading terms are retained, again in the long-distance limit. In order to have a quantitative understanding of this limit, it is useful to calculate
the correlation functions explicitly for the pure Ising model on strips,
and compare their actual rate of decay to that predicted by Eqs. \ref{eq:1} and  \ref{eq:2}. This is illustrated in Fig.~\ \ref{fig:pure}, where two typical 
examples are shown. One can see that, for widely differing temperatures and
strip widths, the influence of subdominant terms on the correlation decay is
always very small and becomes negligible after no more than two lattice spacings. 
We shall invoke this in what follows.     

 For a random system, the decay of thermal correlations for a given realisation
of randomness is still
given through the iterative product of transfer matrices from one spin
column to the next; however, some form
of configurational averaging is now necessary. In previous work~\cite{cheung,huse,glaus1,glaus2}
this was done by generating strips of length $N>>1$ and carrying out the
actual transfer-matrix products, so the end result would presumably
reflect the properties of a representative sample. While the average free energy per site is unambiguously related to the dominant Lyapunov exponent of the  transfer-matrix product\cite{ranmat}, the relationship of correlation functions to Lyapunov exponents is much more involved, because the probability distribution of the correlation functions themselves turns out to be rather complex\cite{crisanti,derr2}. It can be shown that the {\it most probable} correlation decay is given by
the difference between the two largest Lyapunov exponents\cite{crisanti}, which
constitutes a straightforward generalisation of Eq. \ref{eq:2}.
This may not be the same as the {\it average} correlation decay, which
is the experimentally measurable quantity; the distinction between the two becomes especially important when the (effective) interactions fluctuate in sign, as is the case of spin glasses or when random fields are present\cite{crisanti,derr2}. However, when all interactions are ferromagnetic numerical evidence\cite{unpub} indicates that, at least for two-dimensional Ising systems, the two quantities yield essentially the same results apart from
logarithmic corrections. In what follows, we shall assume this to be  the case
also when there are absent sites, all non-zero interactions being positive. 
    
The implementation of the schemes just described runs into the usual problems of judging when the strip is sufficiently long, so that the sample can be taken as representative. For  diluted systems the additional difficulty would arise of how to avoid the effects of disconnections. In early attempts\cite{kimmo} this
was done by introducing a fictitious ``weak '' bond (of strength $10^{-3}$ that of ``normal'' ones) , linking pairs of sites of which at least one was absent .
Numerical results were, however, unsatisfactory.

Here, we perform the averages in a different way. Our starting point
is the transfer-matrix formulation of the percolation
problem~\cite{derrida}. In this case, which is the
zero-temperature limit of diluted magnets, the  (geometric)
correlation length is correctly given by the largest eigenvalue of a
transfer matrix, whose elements are related to the probability that
two adjacent site columns (each with occupied and vacant sites)
are linked to each other and to the origin. Note that disconnections are
altogether avoided from the start, as only column configurations where at
least one occupied site is part of the infinite cluster\cite{derrida} are considered.
Unconnected configurations can be discarded, as they just account for 
conservation of total probability. Were they to be included, the matrix would
have 1 as its largest eigenvalue, with the second largest eigenvalue giving
the correlation length, in exact correspondence with 
Eq.~\ref{eq:2}~\cite{earlyd}
Care is thus taken of the correlations between frozen spins in the dilute
spin system at absolute zero; as the
temperature is raised from zero, each of the possible links
represented by the non-zero elements of the geometric transfer matrix
is weakened by thermal fluctuations. We propose to take these into
account still within the framework of a single matrix, so the exact
column-by-column character of the average of disorder configurations
will be preserved.

Note however that any approach which directly combines a thermal and
disorder averaging (in $e.g.$ multiplying an extended transfer matrix)
will be physically incorrect for quenched disorder problems.

So, we must suitably modify each non-zero element of
the original matrix; obviously we
 must look at the properties of the {\it spin}
transfer matrix between the corresponding site column states. 
As the physical property under investigation
is the rate of decay of correlations,
we draw on an analogy to the leading contribution towards
this in periodic systems: 
we choose to take the ratio of the two largest eigenvalues of the spin
transfer
matrix and multiply the corresponding geometric transfer matrix
element by it. If $i$ and $j$ are two site column states connected
to each other and to the origin, with
respective probabilities $P_{i}$ and $P_{j}$, and the spin transfer
matrix $T^{ij}$ between these columns has $\lambda^{ij}_{1}$ and
$\lambda^{ij}_{2}$ as its largest eigenvalues, the matrix element of
the ``thermal-geometric'' transfer matrix of our formulation is then :
\begin{equation}
{\cal T}_{ij}=\sqrt {P_{i}P_{j}}(\lambda^{ij}_{2}/\lambda^{ij}_{1})\ .
\label{eq:3}  
\end{equation}
\noindent The averaged correlation length in this approximation is given
by the largest eigenvalue, $\Lambda_{1}$, of $\cal T$ via
\begin{equation}
(\xi^{-1})_{ave}=-\ln \Lambda_{1} \ .
\label{eq:4}  
\end{equation}
The following comments are in order:
 
(a) As the temperature 
$T\rightarrow 0$, the two largest eigenvalues of
all thermal transfer matrices become degenerate, and Eq.~\ref{eq:3} shows
that $\cal T$ reduces to the geometric transfer matrix of Ref.~\onlinecite{derrida}, as it should. Thus, the present calculational scheme
is asymptotically correct in the low-temperature regime. For other disordered
systems this must be true as well, provided that one can start from a suitable
transfer-matrix description of the ground-state correlations.

(b) As the concentration $p$ of magnetic sites approaches 1, the
only remaining non-zero element of $\cal T$ is along the diagonal,
connecting two fully occupied columns; denoting by $\lambda^{pure}_{1}$
and $\lambda^{pure}_{2}$ the largest eigenvalues of the corresponding
thermal transfer matrix,
\begin{equation}
\Lambda_{1}=\lambda^{pure}_{2}/\lambda^{pure}_{1},\ \ \ \ p=1  
\label{eq:5}  
\end{equation}
\noindent and the pure system limit is correctly obtained.

(c) As befits quenched problems, disorder and thermal aspects are {\it
not} being averaged together. We represent the effect of thermal
fluctuations on each {\it given} geometric configuration by the ratio of
the respective largest thermal eigenvalues, and the disorder average is 
performed at a later step, in obtaining the largest eigenvalue of
$\cal T$. Thus we make the analogue of the configurational average of
the factor $e^{-1/\xi}$ in the correlation function (related to
its decay between two adjacent columns) and it is indeed the
correlation function which is self-averaging.
 
(d) The procedure outlined contains the approximation
that the contribution given by each thermal transfer matrix is
truncated and replaced by that of its two largest eigenvalues.
Once this has been done, we follow  lines analogous to those 
of Refs.\onlinecite{cheung,huse,glaus2}
for the calculation of an averaged correlation length in a
random system. We are thus (approximately) averaging the two largest 
Lyapunov exponents, and $not$ the correlation functions.
 If all thermal transfer matrices, corresponding to distinct disorder
configurations, commuted with each other (which is not the case)
 our {\it ansatz}  would be
identical to that of Refs.\onlinecite{cheung,huse,glaus2}; note that the 
probabilistic factors included in our matrix elements would
be mirrored in the relative frequency of configurations 
generated by those methods. On the other hand, the influence of
higher-ranking eigenvalues is expected to die out asymptotically,
thus minimising the truncation effects mentioned above. Having in mind the corresponding results for the pure Ising case, portrayed in Fig~\ref{fig:pure}
above, this does not seem too drastic an assumption. 

\medskip

The results for the critical curve are now presented.  

Along the critical line $T_{c}=T_{c}(p)$, the correlation length $\xi$
diverges in the infinite system limit. In phenomenological scaling
between strip widths $L$ and $L^{\prime}$ this is where~\cite{fsgen}
$\xi_{L}/L=\xi_{L^{\prime}}/L^{\prime}$. This condition would normally give
a fixed point in a one-parameter space. Here, where two variables $T$,
$p$ occur, we fix $p$ and find the associated critical temperature
$T_{c}(p)$ for each $(L,L^{\prime})$. As usual, we take periodic boundary conditions along the finite direction of the strips, and $L^{\prime} = L-1$.
Sparse-matrix techniques were used for the diagonalisation of the thermal
transfer matrices. We have obtained the approximate phase diagrams using
phenomenological scaling with $L$ up to 7. For each temperature and concentration at the next step, $L=8$, one would need to take into account nearly 32,000 configurations of connected
pairs of columns and extract the two largest thermal eigenvalues
of the respective transfer matrices; though symmetry considerations can
reduce the number of distinct configurations involved, analysis
of a few selected points convinced us that no additional insight would be gained this way. Data from $L$ = 5, 6 and 7 scalings are shown in Fig.~\ref{fig:tcvsp}, together with an interpolated curve incorporating
the exact values of $T_c(p=1)$ and of the limiting slope at $p=1$, as well as
the best estimate for the percolation threshold $p_c= 0.592745 \pm 0.000002$
\cite{ziff}. The overall picture is similar to those provided by simple
analytic scaling approaches\cite{stinch} and Monte Carlo simulations\cite{brady}. 

 At the extreme points $p=p_c$, $T_c=0$
and $p=1$, $T_c=T_c(p=1)$ it is known (respectively from Refs. \onlinecite{derrida} and \onlinecite{night}) that the approach of strip-scaling
fixed points to the exact critical parameters is monotonic as $L$ grows; as remarked in (a) and (b) above, these sequences  are automatically reobtained in our scheme. For intermediate concentrations, the same uniform convergence does not seem to be present, as curves from successive scalings cross each other. This is almost certainly related to the approximations involved in the truncations referred to above. However, the curves are very close to each other, and to
the interpolated phase boundary. Thus, although the truncation effects
do not die away with increasing $L$ (and there is no {\it a priori} reason they should do so), they are most probably of a quantitatively limited nature.

 We now turn to results for the reduced slope
$\left({1/T_{c}(1)}\right)\left(dT_{c}(p)/dp\right)_{p=1}$ 
of the critical curve at the pure limit, which is known exactly\cite{thmg}.
Close to $p=1$ we were able to reach $L=11$ by using a truncated basis of connected states, 
consisting only of pairs of adjacent columns with at most two vacant sites
overall. We checked the consistency of results
thus obtained against those from the full basis (for $L \geq 6$) and from a basis with up to four vacant sites per two-column configuration (for 
$L \geq 8$).  We made use of analytical expressions linking $dT_{c}(p)/dp$ with
$\partial\xi/\partial T$ and $\partial\xi/\partial p$, through the condition $\xi_{L}/L=\xi_{L^{\prime}}/L^{\prime}$, which defines the critical
line. Both partial derivatives can be evaluated by first-order perturbation
theory instead of numerically; thus, calculations were performed very close to $p=1$ ($1-p=10^{-8}$ was typical), without loss of accuracy . 

       Results from  $(L,L-1)$ scalings up to $L=11$ are given in Table 1,
which complements the corresponding table of Ref. \onlinecite{dqrbs}. The
sudden discontinuity when one goes from  $(L,L^{\prime})$= (7,6) to (8,7) is similar to what takes place for the fixed point of percolation between  $(L,L^{\prime})$ = (4,3) and (5,4)\cite{derrida,earlyd}, where convergence is interrupted.The
extrapolation quoted in Table 1 was obtained by taking the last  four points of the sequence, and searching for the value of  $\psi$ such that the plot of those
data against $L^{-\psi}$ was the best possible straight line (see the work of Derrida and Stauffer in Ref.\onlinecite{derrida}). Here, one has 
$\psi \simeq 2.0$. The exact slope is due to Thorpe and McGurn\cite{thmg}. 
The behaviour of the critical curve near the percolation threshold is
determined by a crossover exponent $\phi$ giving the power-law
dependence of $e^{-2J/T_{c}(p)}$ on $(p-p_{c})$. As $\phi$ is
known exactly to be 1~\cite{coniglio}, this can provide another test of the
quality of the strip scaling results.

At low temperatures, our {\it ansatz} is
asymptotically exact. Care must be taken, however, with the spin degrees of
freedom of occupied sites that are not directly connected to the origin \cite{derrida,earlyd}. With periodic boundary conditions, these may occur for
$L \geq 4$. No matter whether they belong to finite clusters
or are connected by a path that goes forwards before turning backwards to
the origin, such sites do not contribute to the spread of thermal correlations along the backbone of the infinite cluster (they do, on the other hand, matter for the statistics of geometric configurations, so their probabilistic weight must be correctly included when counting the latter). In the corresponding
thermal transfer matrices, the occupied unconnected sites are then treated as if they were absent. This is done, of course, for all temperatures and concentrations, but has well-marked effects in the low-temperature region. We have found that, if unconnected spins are not frozen out, the $T_C \times p$ curves for $L \geq 4$ develop an unphysical overhang at $T \rightarrow 0$, $p < p_c$ before homing in towards the correct $p_c$ at $T=0$. Once these corrections
are incorporated, we get the correct vertical approach to $p_c$ in the $T-p$ plane.

Slow convergence problems put a limit to the lowest temperatures within reach
of investigation; this is usually of order $10\%$ of the critical temperature  
for the pure system, $T_{c}(1)$; for the smallest strip widths, $l$ = 2 -- 4
we get to $\sim 0.06 T_{c}(1)$. As the phase boundary is vertical in that
neighbourhood, the corresponding values of  $p - p_{c}$ are $\sim 10^{-5}$.
This is enough to see the exponential behaviour, with $\phi=1$ along more than two orders of
magnitude of variation in $p - p_{c}$. In Fig. \ref{fig:lt} we show, for $L$= 5,
6 and 7, the quantity    
\begin{equation}
C_{L} \equiv {\exp (-2J/ T_{c}(p)) \over p-p_{c}(L)} \ \ ,  
\label{eq:6}  
\end{equation}
\noindent against $\ln (p-p_{c}(L))$ (where $p_c(L)$ is the fixed point of
phenomenological renormalisation, and $T_{c}(p)$ is obtained from our
{\it ansatz}, both for strips of widths $L$ and $L-1$). For $L$= 6 and 7, the region in which $C_L$ is approximately constant goes from $ p-p_c(L) \simeq 4 \times 10^{-5}$ to $\simeq  6 \times 10^{-3} $. The approximate values at which
the  $C_L$ become stable are: 6.1, 3.8 and 2.3 respectively for $L$= 5, 6 and 7. It does not seem possible to obtain a reasonable estimate of the limiting value of this coefficient, $C_{\infty}$ from extrapolation of this sequence. For
bond dilution, a value for comparison is\cite{domany} $2 \ln 2 = 1.386 \dots$.
A heuristic generalisation for site dilution (see \onlinecite{stinch}, pg. 190) would lead to $C_{\infty}= (1/p_{c})\ln 2 = 1.169 \dots$. The interpolated curve used in
Fig.\ref{fig:tcvsp} has $C \sim 4$, for consistency with corresponding
data of Ref.\onlinecite{dqrbs}; for purposes of overall comparison of
the shape of the phase diagram, the difference relative to the presumed value
$\simeq 1.169$ is not relevant. 

It must be remarked that the region covered in Fig.\ref{fig:lt} is very difficult to reach {\it e.g.} in Monte-Carlo simulations, as one is at 
concentrations extremely close to criticality. For comparison, the lowest
$p$ used in recent work\cite{brady} corresponds to $ p-p_c \simeq 7 \times 10^{-3}$. From the data depicted above, it seems safe to conclude that the
exponential behaviour is dominant at least up to a critical temperature of
order $T_c/J \sim 0.7$.  

Conformal invariance\cite{cardy} allows one to extract additional information from strip scaling, via the relationship
between correlation-length amplitudes on a strip of width $L$
at criticality and the
decay-of-correlations exponent $\eta$:
\begin{equation}
 \eta=L/\pi \xi (T_{c})\ \ .  
\label{eq:7}  
\end{equation}
\noindent In the present case, one has two questions to answer: first, whether conformal invariance still is valid for random systems: second, if so, how to 
define the correlation length which enters Eq.\ref{eq:7}. As for the former,
transfer-matrix\cite{glaus2} and field-theoretical results\cite{ludwig} indicate
the affirmative, provided that one considers averages over disorder. The latter
is more involved, with the two most obvious candidates being the correlation decay factors related to the ``most probable'' and ``average'' correlation functions\cite{ranmat,crisanti}. Based on numerical evidence for the bond-disordered two-dimensional Ising system\cite{unpub}, we assume that for dilution the distinction between results coming from either definition amounts
at most to logarithmic corrections. Thus, we shall use the average correlation
length given by Eq.\ref{eq:4} which, as remarked above, is closely related
to an average of the two largest Lyapunov exponents, i.e., the most probable
correlation decay.     

In Fig.\ref{fig:etall} we display the values of $\eta$ along the approximate
critical curves. As can be seen, behaviour is not uniform with $L$; however,
the overall trend for the exponent is to keep an approximately constant value
close to that of the pure Ising model, $\eta_{I}=1/4$, and drop towards the
percolation value  $\eta_{p}=5/24$ close to $p_c$. 

The above-mentioned trend becomes much more apparent, and uniform with $L$,
when one considers the variation of $\eta$ along the interpolated critical
curve. Fig.\ref{fig:etai} illustrates this. Note the difference between vertical scales in Figs.\ref{fig:etall} and \ref{fig:etai}. Recall from Fig.\ref{fig:tcvsp} that the approximate and interpolated
curves are very close in $T-p$ space. This shows that the averaged correlation length is very sensitive
to slight variations in the parameters.
In both figures, it is apparent that the values of
$\eta$ at $p=p_{c}$, $p=1$ approach the exact values as $L$ increases. Again, the two special cases of the pure Ising model and percolation have been
treated previously~\cite{derrida,night}, and our results for these limits coincide with those already obtained.

The behaviour depicted above is to be compared with current field theory results
for random Ising models in two dimensions~\cite{ludwig,dotsenko,wang}. The theories agree in predicting that the
specific heat singularity is of $\ln\ln(T-T_{c})$ type rather than the
$\ln(T-T_{c})$ of the pure case. But one group of theories~\cite{ludwig}
predicts that the asymptotic pair correlation is as in the pure
case ($<\sigma_{0}\sigma_{r}>~\sim~r^{-\eta}$,~$\eta=1/4$) while the
other~\cite{dotsenko} concludes that the correlation has the form
$<\sigma_{0}\sigma_{r}>~\sim~e^{-A(\ln r)^{2}}$.

Our results clearly support the first class of predictions, and show
no sign of the huge change of $\eta$ expected if the second prediction
were correct. 

Large Monte Carlo simulations on the random-bond Ising
model at a particular (self-dual) concentration~\cite{wang} also provide
indirect evidence in favour of the first class of results. An
evaluation of the susceptibility exponent $\gamma$ is taken together
with scaling relations to infer a value for $\eta$, which is close to
the pure value 1/4. A direct evaluation of $\eta$, again pointing
to the pure result, has also been obtained~\cite{glaus2}
for the same random bond Ising model by strip scaling on long
realisations of the random system. Very recent Monte Carlo results for the correlation functions at criticality also support this view\cite{talapov}.

In conclusion, application of the calculational scheme just described to
the site-diluted Ising model gives reliable and fairly accurate results,
though convergence towards the exact values as strip width $L$ increases
seems to be generally non-uniform. Our data for the exponent $\eta$  at intermediate concentrations give clear evidence in favour of one of the competing classes of field-theoretic treatments\cite{dotsenko,ludwig,wang,talapov}.
 
A straightforward extension of the method to the random-field Ising model in two dimensions gives results for the averaged correlation length  which are
qualitatively similar to those predicted from the two largest Lyapunov 
exponents\cite{dmetc}. These, in turn, seem to differ appreciably from the
correlation lengths obtained directly from direct fits of the averaged correlation decay. As remarked above, cases such as this where frustrations
are present tend to much subtler than when all correlations are ferromagnetic. We defer a detailed analysis of this point to a forthcoming publication.

\acknowledgements 

SLAdQ would like to thank the Department of Theoretical Physics
at Oxford, where this  work was initiated, for the hospitality, and
the cooperation agreement between Academia Brasileira de Ci\^encias and
the Royal Society for funding a later visit to Oxford.
Research of SLAdQ
is partially supported by the Brazilian agencies Minist\'erio da Ci\^encia
e Tecnologia, Conselho Nacional 
de Desenvolvimento Cient\'\i fico e Tecnol\'ogico and Coordena\c c\~ao de
Aperfei\c coamento de Pessoal de Ensino Superior.


\newpage
\vskip 4.0cm
\begin{table}
\caption{
Reduced slope at $p=1$}
\vskip 1.0cm 
 \halign to \hsize{\hskip 5.0cm\hfil#\hfil&\hfil#\hfil\cr
  $L/L'$ & ${1\over T_{c}(1)}\left({dT_{c}(p)\over
 dp}\right)_{p=1}$ \cr
 3/2 & 1.4461 \cr
 4/3 & 1.4765 \cr
 5/4 & 1.5032 \cr
 6/5 & 1.5165 \cr
 7/6 & 1.5223 \cr
 8/7 & 1.4944 \cr
 9/8 & 1.5015 \cr
 10/9 & 1.5065 \cr
 11/10 & 1.5101 \cr
 Extrapolated & 1.53 $\pm$ 0.015 \cr
 Exact & 1.565 \cr}
\end{table}

\begin{figure}
\caption{
Semi-logarithmic plot of correlation function  $<\sigma_0 \sigma_R>$ (squares) against distance $R$ for the pure
Ising model on strips of width $L$ of a square lattice. Values of $\xi$ in lattice spacing units, are as given by Eq. 2.
 Straight lines have slope $ -1/\xi$. $T_c$ is the critical
temperature of the two-dimensional lattice.}
\label{fig:pure}
\end{figure}
\begin{figure}
\caption{
Approximate phase diagram from $(L,L-1)$ scalings for $L$ =5, 6 and 7 and
from interpolation between exact limits.}
\label{fig:tcvsp} 
\end{figure}
\begin{figure}
\caption{ 
Exponential behaviour of $T_c$ against $p$ close to $p_c$. See Eq. 6 
for definition of $C_L$.}
\label{fig:lt} 
\end{figure}
\begin{figure}
\caption{
Correlation exponent $\eta$ from correlation-length
amplitudes, along approximate ($L,L-1$) critical curves. Black squares are at (0.59275,~5/24) and (1,~1/4) .}
\label{fig:etall} 
\end{figure}
\begin{figure}
\caption{
Correlation exponent $\eta$ from correlation-length
amplitudes, along interpolated critical curve. Black squares are at (0.59275,~5/24) and (1,~1/4).}
\label{fig:etai}  
\end{figure}


\begin{references}
\bibitem[*]{email}
Electronic address: sldq@fis.puc-rio.br
\bibitem{baxter}
See for example R.J. Baxter, {\it Exactly Solved Models
in Statistical Mechanics} (Academic, New York, 1982).
\bibitem{fisher}
M.E. Fisher, in {\it Proceedings of the ``Enrico Fermi'' International School of Physics, Varenna, 1970, Course No. 51}
edited by M.S.Green (Academic, New York, 1971).
\bibitem{fsgen}
M.N. Barber, in {\it Phase Transitions and Critical Phenomena} Vol.
8, edited by C. Domb and J.L. Lebowitz (Academic, New York, 1983);
M.P. Nightingale, in {\it Finite Size Scaling and Numerical Simulations
of Statistical Systems}, edited by V. Privman (World Scientific,
Singapore, 1990).
\bibitem{dersaw}
B. Derrida, J. Phys {\bf A14}, L5 (1981).
\bibitem{malte}
M. Henkel, in {\it Finite Size Scaling and Numerical
Simulations of Statistical Systems}, edited by V. Privman (World
Scientific, Singapore, 1990) .
\bibitem{cheung}
H.-F. Cheung and W.L. McMillan, J. Phys. {\bf C16}, 7027
(1983); {\it ibid.} {\bf C16}, 7033 (1983).
\bibitem{huse}
D.A. Huse and I.M. Morgenstern, Phys. Rev. {\bf B32}, 3032 (1985).
\bibitem{glaus1}
U. Glaus, Phys. Rev. {\bf B34}, 3203 (1986).
\bibitem{glaus2}
U. Glaus, J. Phys. {\bf A20}, L595 (1987).
\bibitem{dqrbs}
S. L. A. de Queiroz and R. B. Stinchcombe, Phys. Rev. {\bf B46}, 6635 (1992).
\bibitem{stinch}
See for example R.B. Stinchcombe, in {\it Phase
Transitions and Critical Phenomena} Vol. 7, edited by C. Domb and
J.L. Lebowitz (Academic, New York, 1983). 
\bibitem{derrida}
B. Derrida and L. de Seze, J. Phys. (Paris) {\bf 43}, 475 (1982);
B. Derrida and D. Stauffer, {\it ibid.} {\bf 46}, 1623 (1985) .
\bibitem{night}
M.P. Nightingale, J. Appl. Phys. {\bf 53}, 7927 (1982).
\bibitem{thmg}
M.F. Thorpe and A. McGurn, Phys. Rev. {\bf B20}, 2142 (1979).
\bibitem{domb}
C. Domb, Adv. Phys. {\bf 9}, 149 (1960) .
\bibitem{ranmat}
A. Crisanti, G. Paladin and A. Vulpiani, {\it Products of
Random Matrices in Statistical Physics} (Springer-Verlag, Berlin, 1991).
\bibitem{crisanti}
A. Crisanti, S. Nicolis, G. Paladin and A. Vulpiani, J. Phys. {\bf A23}, 3083 (1990).
\bibitem{derr2}
B. Derrida,  Phys. Rep. {\bf 103}, 29 (1984).
\bibitem{unpub}
S.L.A. de Queiroz and R.B. Stinchcombe (to be published, 1994)
\bibitem{kimmo}
K.K. Kaski, D. Phil Thesis, University of Oxford (unpublished, 1980).
\bibitem{earlyd}
B. Derrida and J. Vannimenus, J. Phys. Lett. (Paris) {\bf 41}, L-473 (1980).
\bibitem{ziff}
R.M. Ziff and B. Sapoval, J. Phys. {\bf A19}, L1169 (1986) .
\bibitem{coniglio}
A. Coniglio, Phys.Rev.Lett. {\bf 46}, 250 (1981)
\bibitem{domany}
E. Domany, J. Phys. {\bf C11}, L337 (1978)
\bibitem{brady}
A.J.F. de Souza and F.G. Brady Moreira, Europhys. Lett. {\bf 17}, 491 (1992)
\bibitem{cardy}
J. L. Cardy, in $Phase\ Transitions\ and\ Critical\ Phenomena$, edited by C. Domb and J.L. Lebowitz (Academic, New York, 1987), vol. 11 . 
\bibitem{ludwig}
 A.W.W. Ludwig, Nucl. Phys. {\bf B330}, 639 (1990).
\bibitem{dotsenko}
 Vik.S. Dotsenko and Vl.S. Dotsenko, J. Phys. {\bf C15}, 495 (1982)
\bibitem{wang}
 J.-S. Wang, W. Selke, Vl.S. Dotsenko and  V.B. Andreichenko, Physica 
{\bf A164}, 221 (1990).
\bibitem{talapov}
A.L. Talapov and L.N. Shchur, preprint hep-lat/9404002 (1994)
\bibitem{dmetc}
D. Moore, R.B. Stinchcombe and S.L.A. de Queiroz, work in progress.

\end{references}
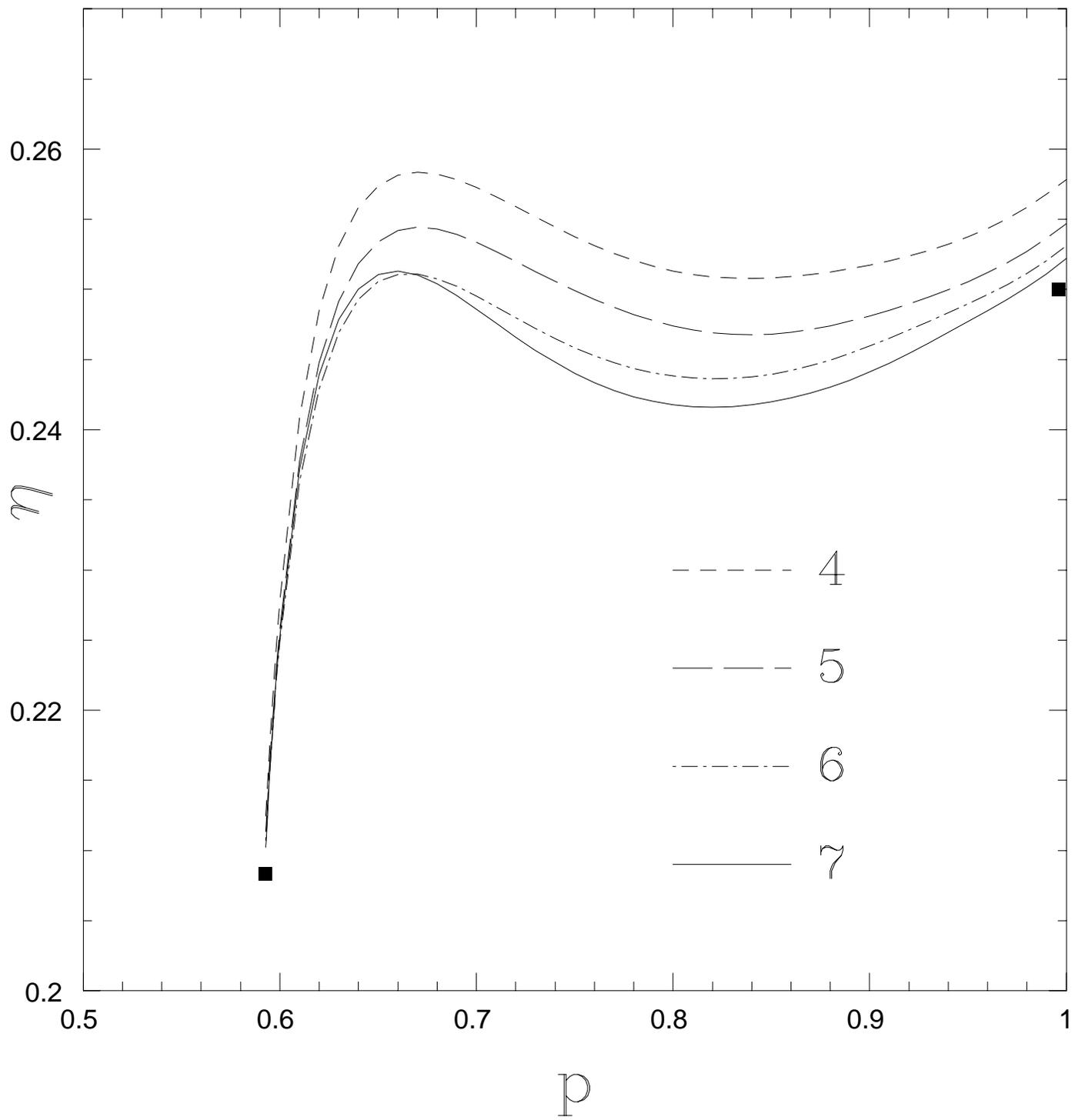
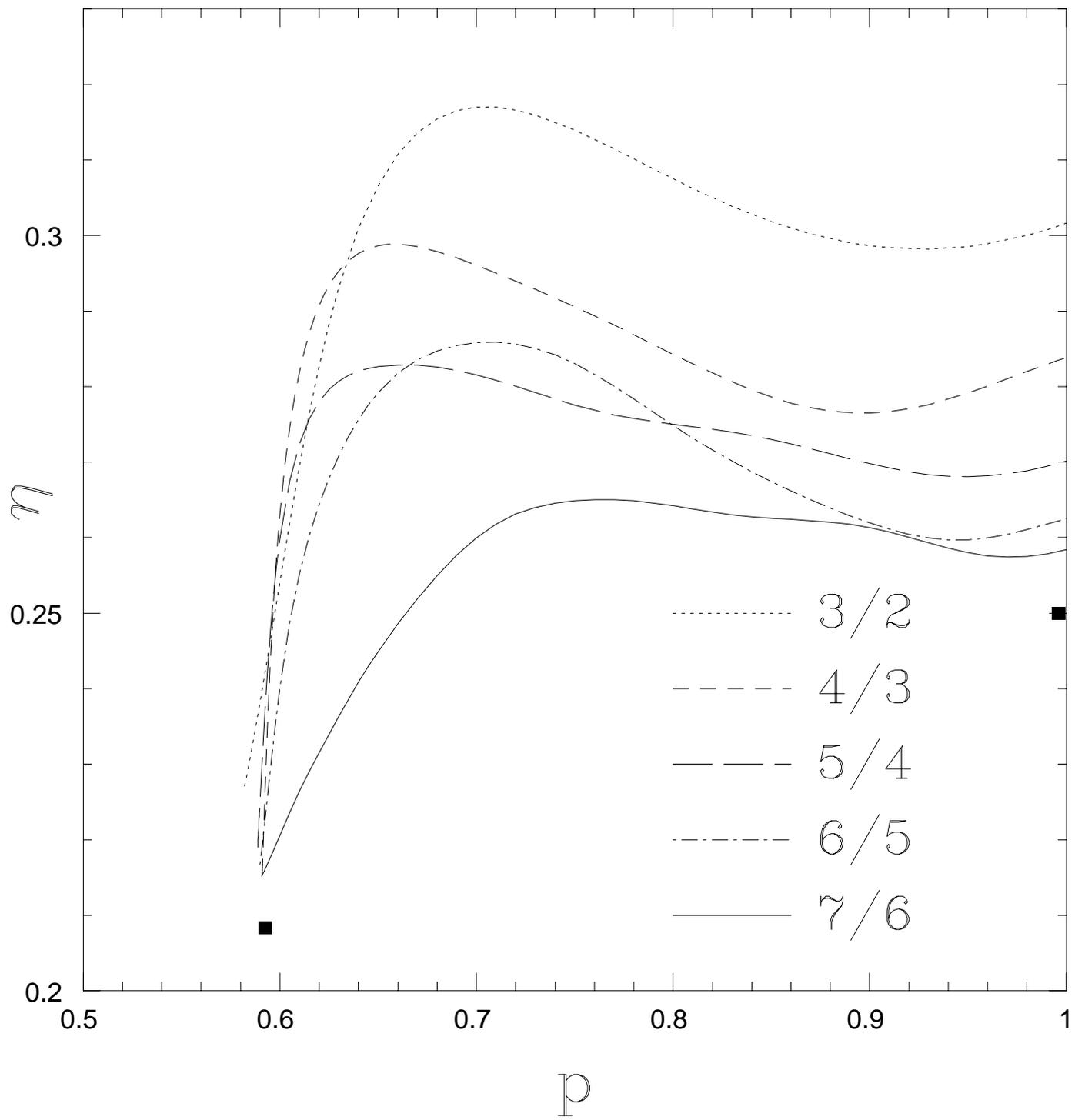
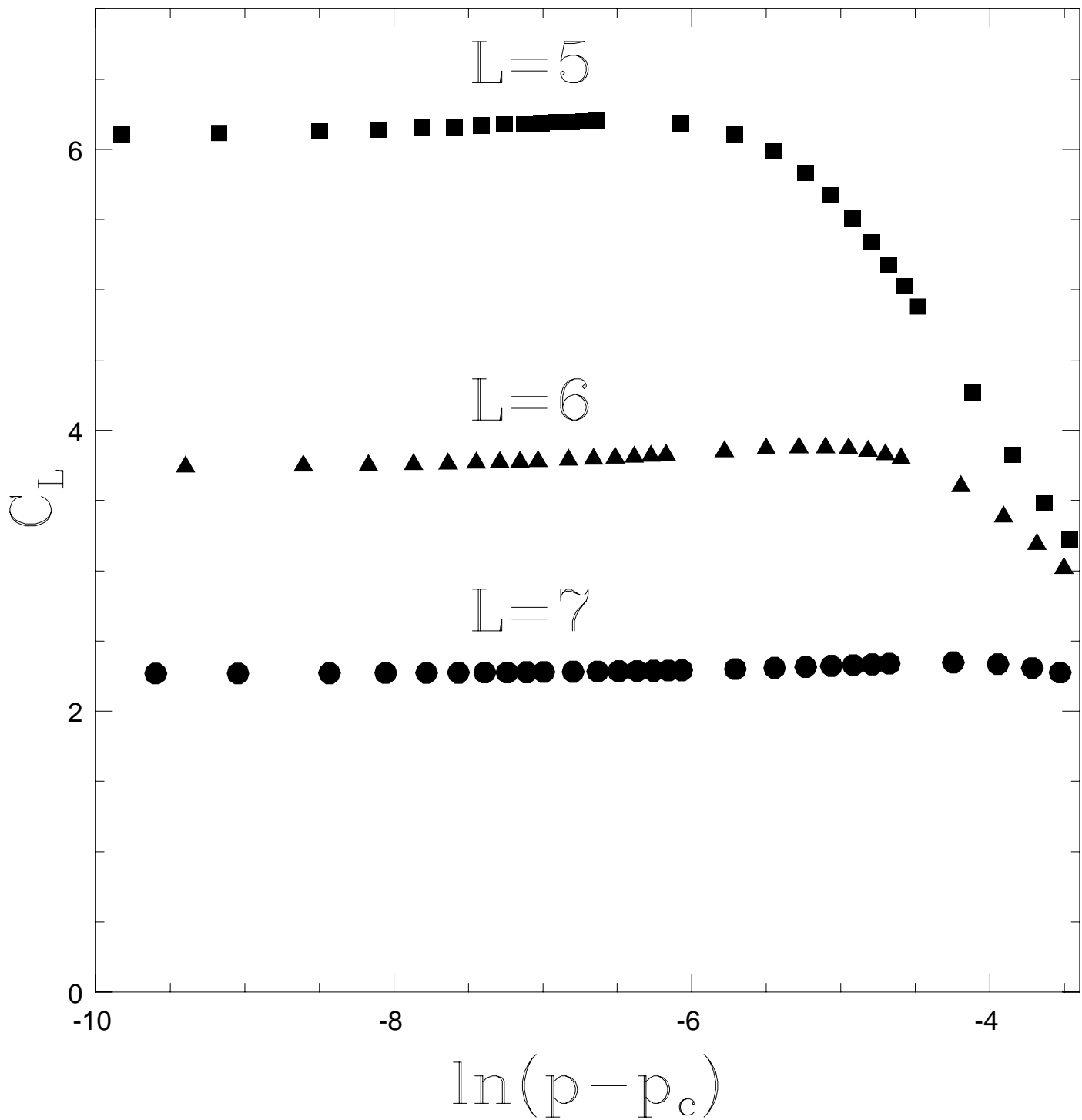
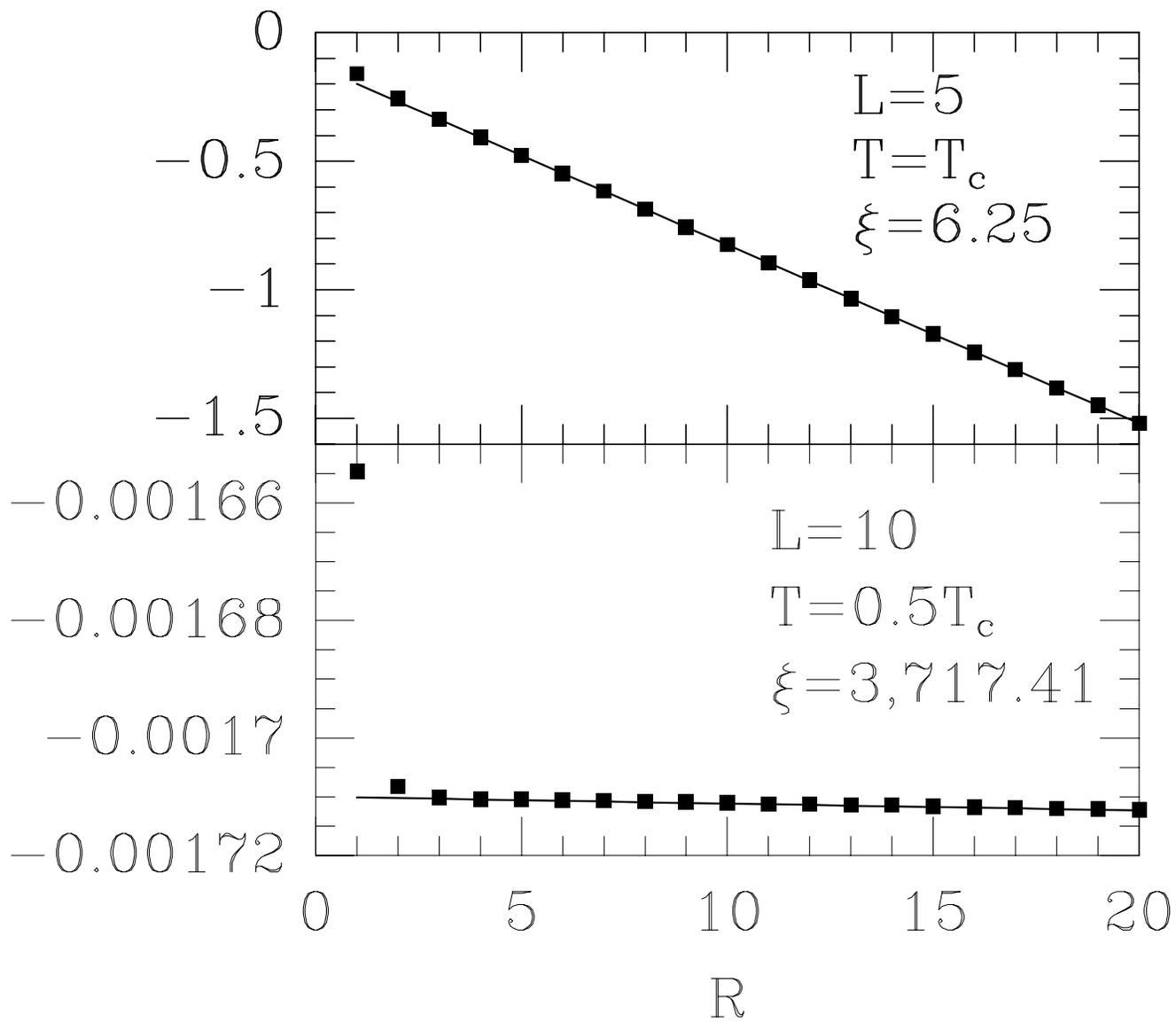
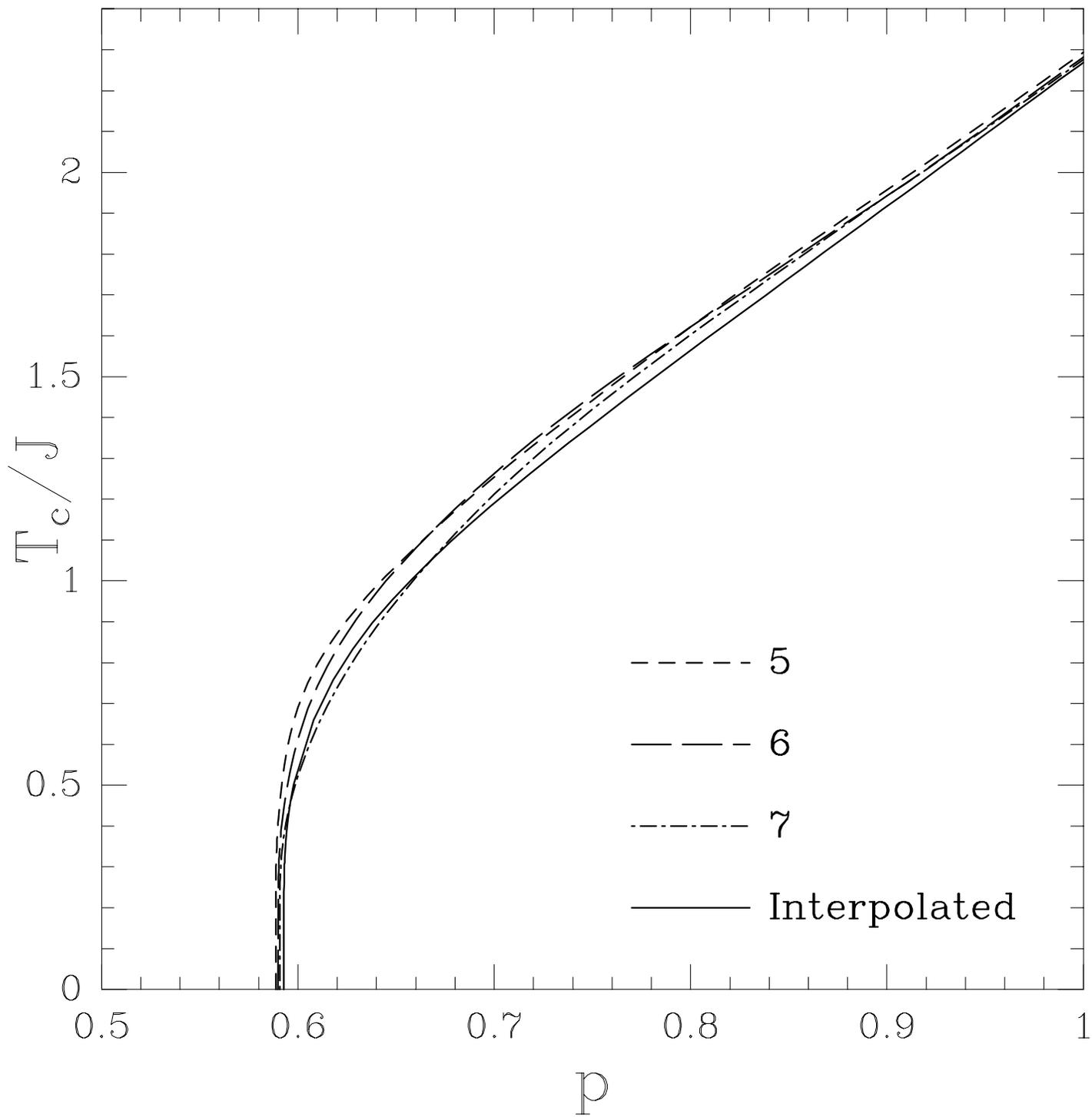
\end{document}